\documentclass[aps,prl,twocolumn,superscriptaddress,showpacs]{revtex4}
\usepackage{amsmath}
\usepackage{dcolumn}
\usepackage{graphicx}
\usepackage{latexsym}
\usepackage{float}

\begin{document}

\title{Confinement-deconfinement transition as an indication of spin liquid type behavior in Na$_2$IrO$_3$}

\author{Zhanybek Alpichshev}
\affiliation{Department of Physics, Massachusetts Institute of Technology, Cambridge, MA 02139}

\author{Fahad Mahmood}
\affiliation{Department of Physics, Massachusetts Institute of Technology, Cambridge, MA 02139}

\author{Gang Cao}
\affiliation{Department of Physics and Astronomy, University of Kentucky, Lexington, KY 40506}

\author{Nuh Gedik}
\affiliation{Department of Physics, Massachusetts Institute of Technology, Cambridge, MA 02139}

\email{gedik@mit.edu}

\date{\today}

\begin{abstract}
We use ultrafast optical spectroscopy to observe binding of charged single-particle excitations (SE) in the magnetically frustrated Mott insulator Na$_2$IrO$_3$. Above the antiferromagnetic ordering temperature ($T_N$) the system response is due to both Hubbard excitons (HE) and their constituent unpaired SE. The SE response becomes strongly suppressed immediately below $T_N$. We argue this increase in binding energy is due to a unique interplay between the frustrated Kitaev and the weak Heisenberg-type ordering term in the Hamiltonian, mediating an effective interaction between the spin-singlet SE. This interaction grows with distance causing the SE to become trapped in the HE, similar to quark confinement inside hadrons. This binding of charged particles, induced by magnetic ordering, is a result of a confinement-deconfinement transition of spin excitations. This observation provides an evidence for spin liquid type behavior which is expected in Na$_2$IrO$_3$.  
\end{abstract}

\pacs{75.10.Kt, 71.10.Li, 78.47.jj}

\maketitle

Spin-orbit coupling (SOC) can give rise to highly non-trivial physics. Prime examples of the role of SOC in condensed matter systems, are the topological insulators which have non-trivial topology of their band structure due to  sufficiently strong SOC \cite{hasan, zhang}. In the case of simpler ``band topological insulators", the gap is determined by the spin-orbit coupling and the system can be treated as non-interacting. These materials have been subject to an intensive research and are relatively well understood. Much less clear is the situation in which the insulating state before ``turning on" of SOC was not a trivial band insulator but an insulator with a gap driven by electron-electron interactions, such as Mott insulator. 

Iridate compounds belong to the class of materials in which electron-electron interactions play an essential role. Expected to be metallic based on simple electron count these systems are insulators exhibiting Mott-type behavior.  On the other hand due to extended nature of the $5d$ orbitals, the on-site Coulomb repulsion has a moderate value ($U\!\approx \, 0.4\! - \!2.5 \, eV$) and SOC ($\approx 0.4\!-\!1 \, eV$) effectively competes with electron-electron interactions \cite{cao}. 

One of the most intriguing proposals of novel physics made for iridate compounds was that the interplay between spin-orbit interactions, crystal field splitting and Coulomb repulsion of $5d$ electrons in Na$_2$IrO$_3$ can lead to a formation of effective moments with $J_{eff}=1/2$ on every Ir-O octahedron with highly anisotropic nearest neighbor coupling. This coupling has a very special form and, given its layered quasi-2D honeycomb lattice structure, was proposed \cite{khaliullin1, khaliullin2, singh1, singh2} to be a solid state realization of the Kitaev model \cite{kitaev} of a spin liquid. 

The real Hamiltonian of Na$_2$IrO$_3$ is however not a pure Kitaev model and should also have conventional terms such as Heisenberg-type exchange interaction between effective moments. Such terms generally spoil the symmetry of the pure Kitaev model and result in an ordered ground state \cite{khaliullin2, singh2, khaliullin3} in the limit of low temperatures. The structure of the ground state should then depend on the details of the extra term in the Hamiltonian. Neutron studies \cite{liu, choi, ye} have revealed that Na$_2$IrO$_3$ has an antiferromagnetic ground state of ``zigzag" type (Fig. 3b) with the N\'eel temperature of $T_N = 15.3 K$. Minimal Hamiltonian within the framework of modified Kitaev model that can give such ground state consists of an antiferromagnetic Kitaev term and a ferromagnetic Heisenberg term \cite{khaliullin2, singh2, khaliullin3, rau}. Although Na$_2$IrO$_3$ is not a quantum spin liquid, the fact that the ordering temperature $T_N=15K$ is considerably smaller than both Curie-Weiss temperature T$_{\Theta}= -125 K$ \cite{singh2} and the spin wave energy $E_{sw} \sim 5 meV$\cite{choi}, implies that the degree of frustration is still quite strong and the Kitaev term should dominate the low energy physics. Nevertheless, despite an intensive research performed on Na$_2$IrO$_3$ so far, to the best of our knowledge, there has been no evidence of the spin liquid type behavior in this material, which should show up in the $T_N\!\ll T\!\ll T_{\Theta}$ temperature range \cite{balents}.

In this Letter we report on our results on ultrafast studies of photo-excitations in  Na$_2$IrO$_3$. We observe a change in their dynamics across N\'eel temperature, namely we observe a sharp increase in the binding energy of the excitons that they form as the system enters ordered phase. We interpret this as an evidence of confinement-deconfinement transition of spin and charge excitations across $T_N$, which is a hallmark of spin-liquid physics \cite{sachdev2, senthil}. 

\begin{figure}[top]
\includegraphics[width=\columnwidth]{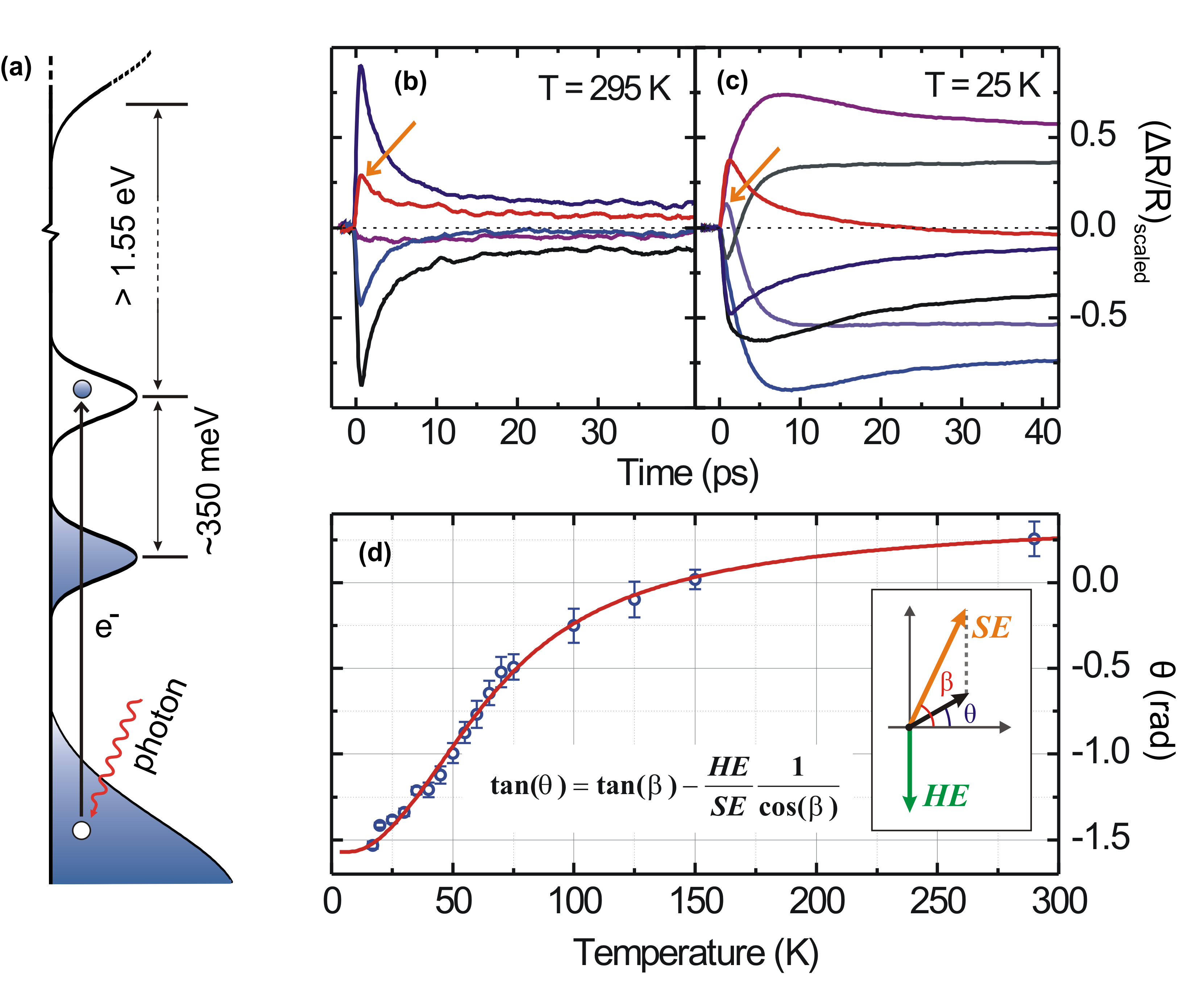}
\caption{(a) Sketch of the band structure of Na$_2$IrO$_3$ illustrating the relevant processes during photo-excitation with $1.5eV$ light (b) HTG traces for selected values of $\phi$ (see text) at a pump fluence of $9.5\mu J/cm^2$ at $T=295 K$ and (c) at $T=25 K$. Note the multi-component behavior of the HTG traces as a function of $\phi$ and the relative strengthening of the initial spike (labeled with an arrow) associated with single-particle excitations (SE) at higher temperature (see text); (d) Phase ($\theta$) of the total signal at $t \approx 50 ps$ in the quasi-equilibrium state as a function of temperature. Error bars represent the $95\%$ confidence interval (2 s.d.) in extracting the phase. Solid red line is a fit to the data based on the Boltzmann distribution of SE and Hubbard exciton (HE) populations, $SE/HE \propto exp(-\Delta/k_BT)$. Inset: phasor diagram representing quasi-equilibrium $\Delta R$ (black phasor) due to SE and HE.}
\label{tg}
\end{figure} 

Time resolved experiments were performed with a Ti:Sapphire oscillator lasing at the center wavelength of $795 \, nm$ ($\hbar \omega = 1.55 eV$) producing pulses of $60 fs$ in duration. The repetition rate of the laser was reduced to $1.6 MHz$ with an external pulse-picker to avoid cumulative heating effects on the sample and the spot size of a spatially Gaussian beam was set to $60 \mu m$ FWHM. Single crystals of Na$_2$IrO$_3$ were grown using a self-flux method from off-stoichiometric quantities of IrO$_2$ and Na$_2$CO$_3$. Similar technical details were described elsewhere \cite{sample1, sample2, sample3}. 

Data was obtained with a standard optical pump-probe technique \cite{pp1, pp2} where a single ``pump" pulse excites the sample and the resulting dynamical response is monitored by the normalized change in the reflectivity $\Delta R(t)/R$ of a separate ``probe" beam as a function of time delay $\Delta t$ between the pump and probe. We use same wavelength for both pump and probe pulses. For phase sensitive measurements a variation of pump-probe technique called the ``heterodyne transient grating" method (HTG) is used \cite{gedik}. This method, unlike pump-probe, can distinguish between components of $\Delta R$ with different physical origins \cite{hinton}. Here an interference of two pump beams produces spatially modulated excitation pattern which is studied by a probe beam diffracted off the sample. The diffracted beam is then heterodyned with an additional beam used as a local oscillator. The time dependence, $\Delta R(t)$, of a multi-component system response obtained this way, changes as a function of the phase difference $\phi$ between the probe beam and the local oscillator, whereas that of a single-component system just scales proportionally to $\cos(\phi)$ (also see Supplementary info).

Given the band structure of Na$_2$IrO$_3$ \cite{rixs, damascelli, sohn} (Fig. 1a), the absorption of a pump photon with energy $E = 1.55 eV$ causes electrons to transition from a $J_{eff}=3/2$ band into the upper Hubbard band (UH) which is the only accessible level for the excited electrons for this photon energy. The depleted valence band is then partially refilled through relaxation of the photo-excited electrons and partially with electrons from the lower Hubbard band (LH) (Fig. 1a). At the end of this relatively fast process the system will have some amount of excitations in the UH (double occupancies in real space or ``doublons'') and an equal amount of holes in the LH. The excited state is metastable as the optical dipole transitions within the Hubbard band are prohibited by selection rules ($\Delta J=0$ transition). Moreover, the energy of magnons which are the relevant excitations $\epsilon \approx 5 - 10meV \sim k_B T_\Theta$ \cite{choi} is much less than the Hubbard gap $U\sim 350 meV$, which is the energy needed to be dissipated during the doublon-hole recombination process, making the lifetime exponentially large in $U/\epsilon$ \cite{demler}. This observation allows us to consider holes and doublons as stable quasi-particles for the timescales relevant to our experiments ($\sim 100 ps$). 

Figures 1b and 1c show HTG data taken at a pump fluence of $\sim 9.5 \mu J/cm^2$ for various values of $\phi$ at $295 K$ and $25 K$ respectively. At each temperature the shape of the differential reflectivity timetrace $\Delta R(t)/R$ with time changes as $\phi$ is varied indicating that there is more than one component in the system response. This behavior is in agreement with earlier Resonant Inelastic X-ray Scattering (RIXS) studies \cite{rixs, gretarsson} which demonstrated that the low energy excitations of Na$_2$IrO$_3$ are single particle (doublons and holes) excitations (SE) and their bound state known as a Hubbard exciton (HE) \cite{parmigiani}. Figures 3a and 3b demonstrate that the component featuring the fast spike near $t=0$ is clearly getting stronger with increasing temperature, allowing us to identify it with SE. This is similar with the results of studies of photo-excited Mott insulators in other systems \cite{parmigiani} where the $\Delta R/R$ component with an initial fast spike was shown to be due to SE whereas the one without due to HE. 

It should be noted that both components (SE and HE) are long lived and thus they both contribute to the composition of the total signal in the long-time limit. There a quasi-thermal equilibrium is established between SE and HE and thus their population ratio should be proportional to the Boltzmann factor $\exp(-\Delta / k_B T)$ where $\Delta$ is the HE binding energy. In this regime, the net phase $\theta$ of the signal response reaches a constant value which is directly related to the quasi-equilibrium population ratio of SE and HE (see Fig. 1d inset and supplementary). This phase $\theta$ is plotted in Fig. 1d as a function of temperature from which we extract $\Delta \approx 4.6 \pm 0.8 meV$. This value is within the bounds set by RIXS measurements \cite{rixs} and confirms that the component featuring a fast spike at $t=0$ is indeed due to SE. 
Fig. 2a and 2b show the HTG data taken for a very low pump fluence of value of $\sim 30 nJ/cm^2$ at temperatures above and below $T_N$ respectively. Similar to the higher fluence data, the low fluence response above $T_N$ (Fig. 2a) clearly features more than one component indicating the presence of both SE and HE. On the other hand, the low fluence response below $T_N$ (Fig. 2b), strikingly, scales proportionally to $\cos(\phi)$ implying a single component behavior which, as discussed above, is due to Hubbard excitons. 

We now proceed to study this disappearance of SE as a function of temperature by performing optical pump-probe measurements for low excitation densities to minimize heating effects. Fig. 2d shows reflectivity transients for various temperatures below $T_N$ for two different pump fluences. As can be seen, the normalized system response in this regime is independent of both temperature and pump fluence demonstrating that the single component behavior observed at $T=5 K$ persists up to $T_N$. This indicates that SE are suppressed throughout the ordered phase. On the contrary, reflectivity transients above $T_N$ (Fig. 2c) strongly depend on temperature, which combined with the HTG data (Fig. 1), indicate the formation and strengthening of the component due to SE. Moreover, above $T_N$, the normalized transients at each temperature are independent of pump fluence (Fig. 2c) demonstrating that the relative composition of the signal (ratio between SE and HE populations) is constant as a function of pump fluence in this low excitation density regime. This sudden disappearance of SE at $T_N$ implies a sharp increase in the HE binding energy.

\begin{figure}[top]
\includegraphics[width=\columnwidth]{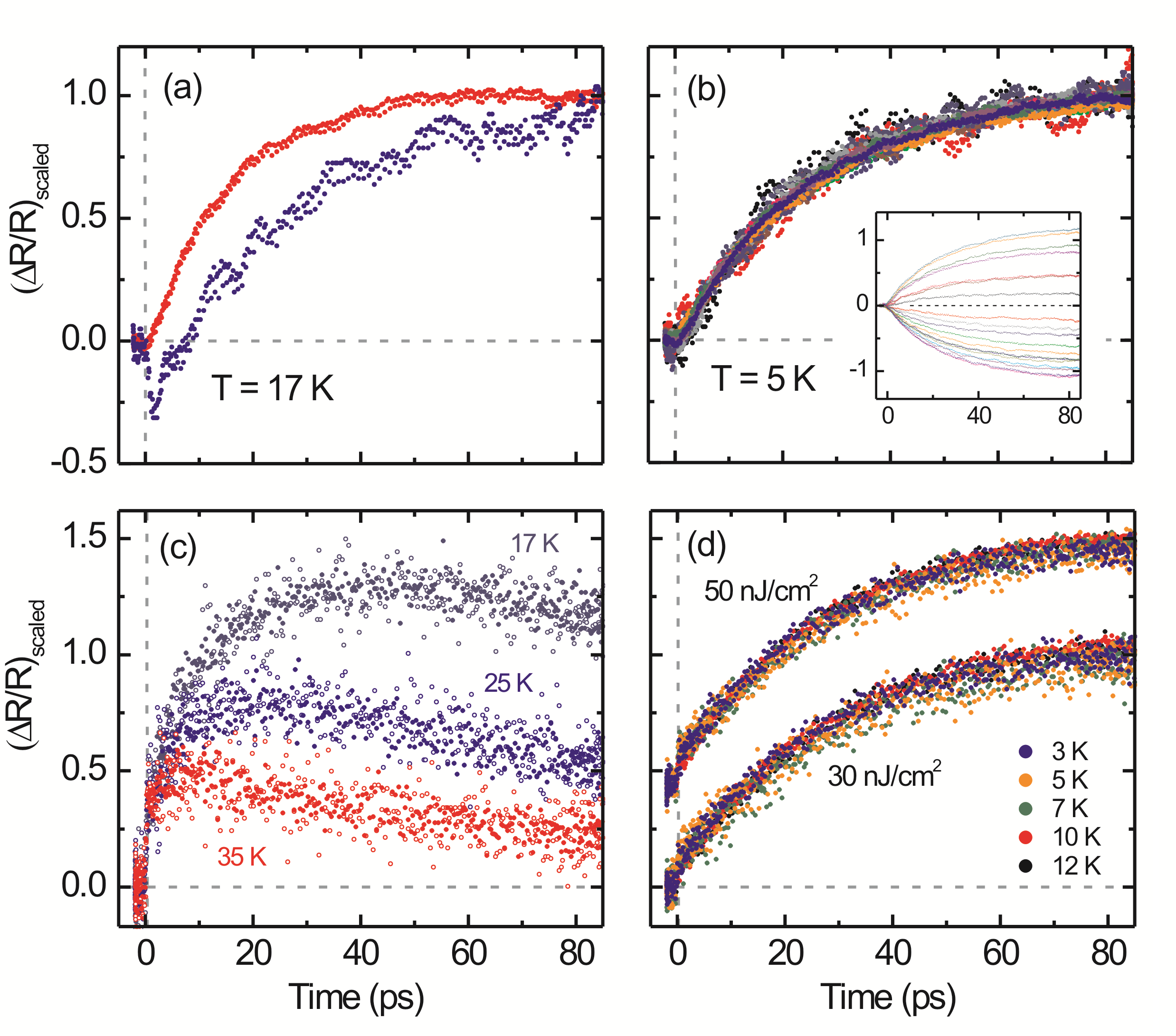}
\caption{\textbf{HTG and pump-probe data at various temperatures at low pump fluence.} a) Representative HTG traces at $T = 17 K$ for two different values of $\phi$ at a pump fluence of $30 nJ/cm^2$. The traces exhibit qualitatively different shaped indicating the presence of both SE and HE. Note the fast spike around $t = 0$ for the purple curve. This is signature of SE (see Fig. 1 and text); b) HTG traces at $T = 5 K$ for 19 different values of $\phi$ at a pump fluence of $30 nJ/cm^2$, scaled to emphasize the single component (HE) nature of the response; inset: unscaled HTG traces; c) Scaled $\Delta R/R$ traces for temperatures above $T_N$ ($T = 17 K$, $25 K$ and $35 K$) at different fluences: $100 nJ/cm^2$ (filled markers) and $50 nJ/cm^2$ (open markers). Note the strong temperature dependence and the lack of fluence dependence in this limit; d) $\Delta R/R$ traces for temperatures below $T_N$ ($T = 5 K$, $7 K$, $10 K$ and $12 K$), scaled to emphasize the universal behavior of transient traces. Upper curve: $50 nJ/cm^2$, lower curve: $30 nJ/cm^2$. Curves at different fluence values are shifted for better clarity.}
\label{pp}
\end{figure}

In general an increase in binding energy of an exciton can be either due to an enhancement of the attracting potential or due to ``slowing down" of the overall dynamics, \textit{e.g.}\! by increasing the effective mass. Effective mass of a single hole in a Mott insulator is indeed enhanced due to emission of magnons, but this happens in both the antiferromagnetic \textit{and} disordered phases \cite{kane}. But more importantly, it is the sheer disparity between the bandwidth of the single particle excitations ($\gtrsim 100meV$, \textit{e.g.} \cite{rixs}) and the Heisenberg coupling responsible for antiferromagnetism ($\sim k_B T_N \approx 1meV$) that makes the ``slowing down" scenario unlikely. Therefore we conclude that the increase in binding energy of HE is a result of enhancement in attraction between SEs.  

\begin{figure}[top]
\includegraphics[width=\columnwidth]{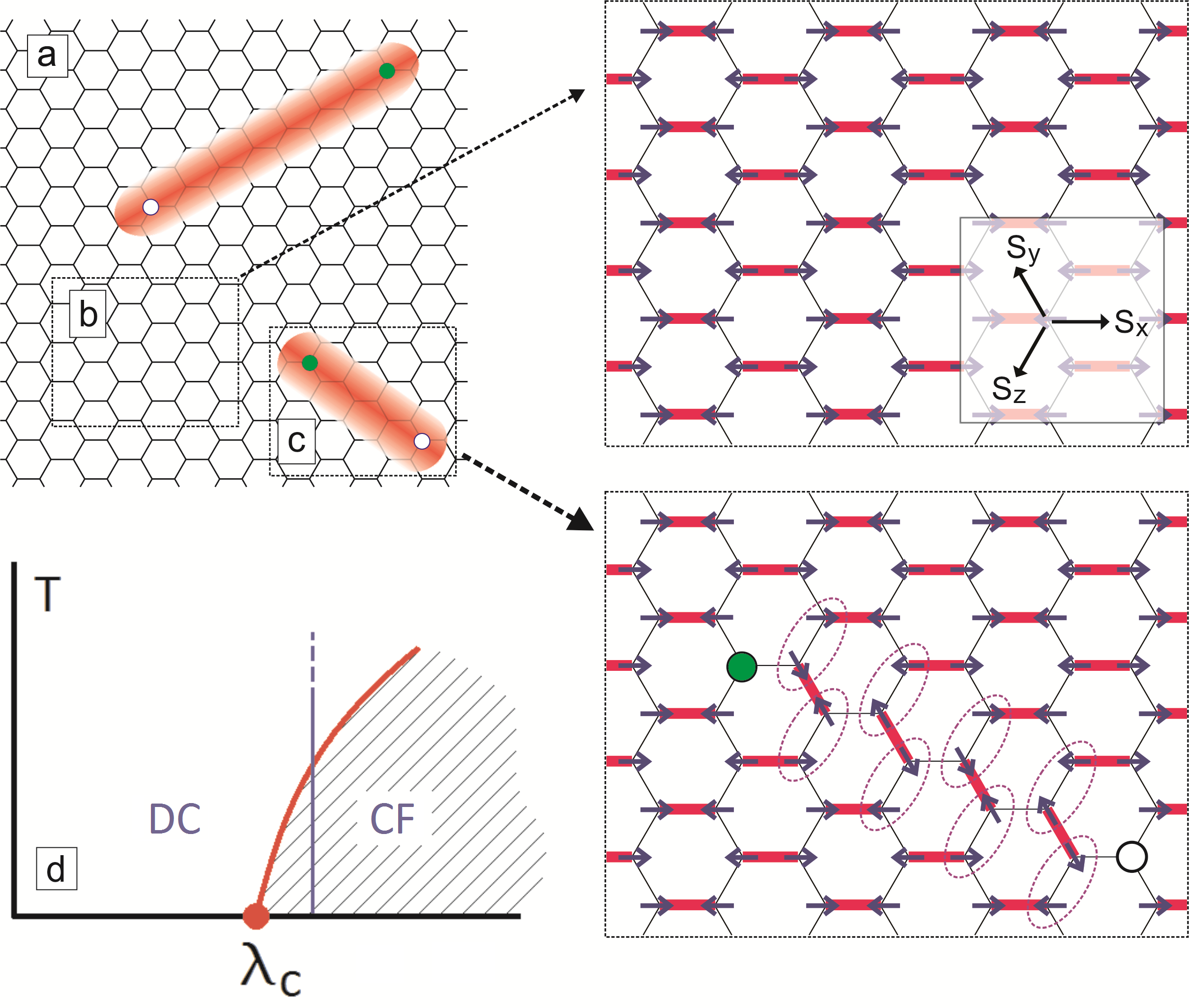}
\caption{a) Schematic view of an excited sample. Green dots correspond to doublons while the white dots represent holes. The shaded red area marks the region where the spins are affected by the reconstruction i.e. the ``string" (see (c)); b) Sketch representing the zigzag ordered low temperature ground state. Red bonds connect Kitaev partners while grey bonds have a spin configuration that minimizes the Heisenberg energy. c) Simplified representation of the restructured state with singlet defects (hole or doublon). The Heisenberg energy along the bonds highlighted by purple ovals is not minimal anymore. The string composed of such bonds must begin and end on a defect. d) Cartoon of a phase diagram of modified Kitaev model with a first-order quantum critical point and a phase boundary between confined (CF) and deconfined (DC).}
\label{order}
\end{figure}   
  
Here we present a simple intuitive picture of the effective attraction mediated by antiferromagnetic ordering responsible for additional binding energy, based on the Kitaev-Heisenberg model. This Hamiltonian naturally gives rise to the zigzag order \cite{khaliullin3}: 

$$H_{KH} = J_{K} \sum \limits_{\langle i j \rangle} S^l _i S^l_j + J_H \sum \limits_{\langle i j \rangle} \vec{S}_i \cdot \vec{S}_j$$ 

\noindent The main term here is the strongly frustrated Kitaev term ($T_{\Theta} = -125 K$). The Heisenberg term lifts the degeneracy and the system ``freezes'' into an ordered state below $T_N$. In the zigzag phase every spin finds its ``Kitaev partner" and anti-aligns itself with the spin of its partner in the direction determined by the orientation of the connecting bond. The much smaller Heisenberg term ($T_N = 15.3 K$) tries to minimize its energy under the condition that every spin has a Kitaev partner. As can be seen in Fig. 3b, zigzag order satisfies this condition for all bonds except those that connect Kitaev partners.

This situation changes drastically when spinless defects such as doublons or holes are introduced into the system. These particles can be thought of as topological in the same sense as the excitations in the Rokhsar-Kivelson dimer model \cite{rokhsar}. The spins that have lost their Kitaev partners reorganize the surrounding spin order at the expense of Heisenberg energy (Fig. 3c). Re-oriented spins form a string terminating on the other defect. The energy cost of this configuration is proportional to the number of broken Heisenberg links which in turn is proportional to the length of the string connecting the two defects. This prohibits the long range separation of the defects which, in the case of low excitation density, will predominantly be of opposite charges (doublon-hole), leading to an enhanced binding between them in addition to Coulomb attraction. This is similar to the picture of the quark confinement in high energy physics \cite{polyakov}: separation of defects produces a string of perturbed vacuum between them with an energy proportional to its length. However, since breaking of this string in our case produces a pair of electrically neutral unpaired ``dangling'' spins (which \textit{are} confined), the binding energy between doublons and holes is limited by the cost of breaking a Kitaev pair, which is of the order of Kitaev coupling $J_K\approx 10meV$ in case of Na$_2$IrO$_3$.

This is consistent with previous theoretical works on Kitaev-Heisenberg model where it was observed that for sufficiently weak perturbations the Kitaev spin liquid state persists but as the extra term gets stronger the system enters an ordered state \cite{khaliullin1, khaliullin2, ybkim}. Formulated in terms of spinons such transtition corresponds to the transition from a deconfined state (spin liquid) to a confined state (antiferromagnet) \cite{baskaran, ybkim}. This is a first order phase transition \cite{baskaran, ybkim} and no quantum critical region is expected above the quantum critical point $\lambda = \lambda_c$ \cite{sachdev3}, where $\lambda$ is the relative strength of the Heisenberg term in the Hamiltonian. Therefore confined and deconfined phases are separated by a simple boundary and the confinement-deconfinement transition can be observed not only by tuning the strength of the perturbing term $\lambda$ at $T=0$ as in Refs.\cite{baskaran, ybkim}, but also by going across ordering temperature for a fixed $\lambda > \lambda_c$ (see Fig.3d). In the confined phase, all fractional excitations such as holons, doublons and spinons are bound to each other, and conversely, can move independently in the deconfined phase \cite{senthil}. 

The fact that the zigzag order is not a trivial antiferromagnet was evidenced in a recent work of Manni \textit{et al.} \cite{gegenwart} where partial substitution of Ir atoms with nonmagnetic Ti atoms resulted in the formation of a spin glass state at low temperatures. This has a natural explanation within our framework, as Ti sites can be treated as static spinless defects. At sufficient concentration they will be connected by spin strings as in Fig.2c. Since there are many different ways to connect different Ti sites, it is natural to expect a spin glass state at low temperatures.

In conclusion, we performed an optical pump probe study of Na$_2$IrO$_3$, a material proposed to be a realization of the Kitaev model. We observed that photo-induced charged excitations display drastically different behavior below and above N\'eel ordering temperature. Namely the binding energy of the excitons that these particles form undergoes a sharp increase upon entering the ordered phase. Based on earlier theoretical studies on doped Mott insulators, we conjecture that this is due to an effective attraction brought about by the antiferromagnetic order rather than because of increase in effective mass of quasiparticles. We argue that this attraction is a manifestation of confinement of spin excitations anticipated in the ordered phase of Kitaev-Heisenberg model. Therefore the change of behavior that we observe at the N\'eel temperature is due to confinement-deconfinement transition, providing an evidence of spin liquid type physics in Na$_2$IrO$_3$.

\acknowledgments{
The authors would like to thank Senthil Todadri, Patrick Lee, Subir Sachdev and Maksym Serbyn for insightful discussions. This work was supported by the Army Research Office Grant No. W911NF-11-1-0331 (data taking and analysis), NSF Career Award DMR-0845296 (experimental setup) 
and by the Alfred P. Sloan Foundation (theory and modelling). GC was supported by the NSF by grants DMR-0856234 and DMR-1265162 (material growth).}

\end{document}